\documentclass[twocolumn,showpacs,preprintnumbers,amsmath,amssymb,superscriptaddress,longbibliography,floatfix,prl,aps,10pt]{revtex4-2}
\usepackage{graphicx}
\usepackage{epsfig}
\usepackage{epstopdf}
\usepackage{dcolumn}
\usepackage{bm}
\usepackage{amsmath}
\usepackage{color}

\usepackage{enumitem}
\usepackage{xspace}
\usepackage{lipsum}

\usepackage{orcidlink}
\usepackage{textcomp}

\usepackage[utf8]{inputenc}
\usepackage[T1]{fontenc}
\usepackage{indentfirst}
\usepackage{textgreek}
\usepackage{multirow}
\usepackage{footmisc}

\makeatletter
\def\maketitle{
\@author@finish
\title@column\titleblock@produce
\suppressfloats[t]}
\makeatother

\newcommand{\disregard}[1]{}

\newcommand{\rmd}{\mbox{d}}

\newcommand{\bfr}{\bm{r}}

\newcommand{\intsum}{\int\hspace{-1.4em}\sum}
\newcommand{\intsumtext}{\int\hspace{-1.0em}\sum}

\newcommand{\Rho}{R}
\renewcommand{\Rho}{\rho}

\newcommand{\trho}{{^2\rho}}
\renewcommand{\trho}{{\cal R}}
\renewcommand{\trho}{{\mathbb{R}}}

\newcommand{\htrho}{{^2\hat{\rho}}}
\renewcommand{\htrho}{{\hat{{\cal R}}}}
\renewcommand{\htrho}{{\hat{{\mathbb{R}}}}}

\newcommand{\tprob}{{^2n}}
\renewcommand{\tprob}{{W}}
\renewcommand{\tprob}{{\mathbb{W}}}

\newcommand{\tcond}{{\mathbb{C}}}

\newcommand{\oprob}{{n}}

\newcommand{\BAR}{\widetilde}

\newcommand{\Kbis}{L}
\newcommand{\LAMBDA}{J''}
\newcommand{\MU}{M''}
\newcommand{\NU}{K''}

\newcommand{\mytitle}{Multipole tomography of atomic nuclei with symmetry-conserved theories}

\newcommand{\myabstract}{To define the intrinsic reference frame and multipole moments of angular-momentum-$J$-conserving many-body wave functions, we introduce two-body conditional probabilities of finding two nucleons at different positions in space. In this way, quadrupole deformations of states with $J\leq1/2$, which are not accessible via spectroscopic one-body quadrupole moments, can be characterized. We illustrate the method with nuclear density functional theory calculations for $J=0$ states of $^{16}$O and $^{20}$Ne, the latter obtained by restoring rotational symmetry of prolate or oblate intrinsic configurations. We show that the two-body quadrupole shape characterizations differ from one-body moments obtained from broken-symmetry states.
}

\begin{document}

\preprint{}

\title{\protect\mytitle}

\title{\mytitle}

\author{X. Sun\orcidlink{0000-0002-0130-6269}}
\affiliation{Department of Physics, University of York, Heslington, York YO10 5DD, United Kingdom}

\author{J. Dobaczewski\orcidlink{0000-0002-4158-3770}}
\affiliation{Department of Physics, University of York, Heslington, York YO10 5DD, United Kingdom}
\affiliation{Institute of Theoretical Physics, Faculty of Physics, University of Warsaw, ul. Pasteura 5, PL-02-093 Warsaw, Poland}

\author{W. Nazarewicz\,\orcidlink{0000-0002-8084-7425}}
\affiliation{Facility for Rare Isotope Beams, Michigan State University, East Lansing, Michigan 48824, USA}
\affiliation{Department of Physics and Astronomy, Michigan State University, East Lansing, Michigan 48824, USA}

\author{H. Wibowo\orcidlink{0000-0003-4093-0600}}
\affiliation{Department of Physics, University of York, Heslington, York YO10 5DD, United Kingdom}

\date{\today}

\begin{abstract}
\protect\myabstract
\end{abstract}

\maketitle

{\it Introduction}---What is the shape of a composite quantum object? Our classical minds seek to answer this question, whereas quantum reality delegates the task to the design of specific observables associated with the notion of shape. Classically, a request for the shape is known in the literature ({\it S'il vous plait \ldots Dessine-moi un mouton}~\cite{(Exu96)}). Quantally, nuclear shapes and observables that characterize them must be extracted from the many-body wave function.

Nuclear shapes and deformations can be described in many ways. In models and theories employing a body-fixed reference frame for the nucleus, such as mean-field approaches, non-spherical shapes arise from the polarization of the nucleonic density by valence nucleons. The resulting nuclear Jahn-Teller effect~\cite{Reinhard1984,Nazarewicz1994} yields intrinsic states that are not eigenstates of operators characterizing the symmetry group of the total Hamiltonian. That is, in intrinsic states, nuclear symmetries can be spontaneously broken~\cite{Anderson72}.

In models and theories explicitly constructed in the laboratory frame, based on symmetry-invariant many-body Hamiltonians, the notion of non-spherical shapes does not appear explicitly. For instance, within the nuclear shell model with a rotationally invariant interaction, all even-even nuclei in their $J^\pi=0^+$ ground states are predicted to have spherical nucleonic densities. To analyze collective excitations in the geometrical language of broken symmetries, it is, however, often instructive in such models to define the intrinsic reference frame. Likewise, the symmetries broken in mean-field models can be restored by projection techniques~\cite{(She21a)}.

For $J\geq1$ states, the nuclear shapes can be characterized by calculating and measuring the spectroscopic average values of one-body quadrupole moments $\hat{Q}_{2\mu}=\sum_{i=1}^A{Q}_{2\mu}(\bfr_i)$~\footnote{Traditionally, the single-particle multipole moments, which in spherical coordinates $\left\{r\theta\varphi\right\}$ are defined through the standard spherical harmonics as $Q_{\lambda\mu}(\bfr)=c_{\lambda}{}r^\lambda{}Y_{\lambda\mu}(\theta,\varphi)$, include suitably chosen numerical prefactors $c_\lambda$, such as $c_2=\sqrt{16\pi/5}$ used here.}. The body of experimental and theoretical data obtained in this way is overwhelming; see, e.g., Refs.~\cite{(Sto21),Dobaczewski2025,(Dob25k)}. However, quantum-mechanical selection rules set all such moments to zero for $J\leq1/2$ states, which includes the ground states of all even-even nuclei. This Letter introduces observables that can replace the spectroscopic moments.

The definition of the intrinsic system depends on the theoretical framework used.
For instance, in nuclear lattice effective field theory (NLEFT)~\cite{Lee2009}, calculations are performed in a 3D spatial box that defines the intrinsic system. The pinhole algorithm~\cite{Elhatisari2017} is used to compute density correlations relative to the center of mass. In this way, the positions of all $A$ nucleons are determined. For a given state with good angular momentum, the wave function is approximately projected onto an irreducible representation of the cubic group. This approach enables exploration of various cluster configurations in light nuclei~\cite{Shen2023}.

The Monte Carlo Shell Model~\cite{Otsuka2022} has also been used to extract nuclear deformations and cluster structures by defining the intrinsic density in terms of the aligned state, in which each basis state in the many-body wave function is rotated to align with its principal axes. However, as demonstrated in the \href{https://static-content.springer.com/esm/art%3A10.1038%2Fs41467-023-38391-y/MediaObjects/41467_2023_38391_MOESM2_ESM.pdf}{Peer Review File} of Ref.~\cite{Shen2023},  the definition of such an aligned state is not unique, and ``nearly any final result can be obtained by choosing different basis states.''

The many-body trial wave function of fermionic molecular dynamics~\cite{Roth2004} is the Slater determinant of localized Gaussian wave packets. The resulting intrinsic one-body densities can be deformed~\cite{Roth2004,Chernykh2007}; hence, angular momentum must be restored by projection. A similar strategy has been applied in the discrete non-orthogonal shell model~\cite{Dao2022} and projected generator coordinate method~\cite{Bally2019} models, in which the wave function is spanned over projected  Slater determinants representing Hartree-Fock (HF) or Hartree-Fock-Bogoliubov (HFB) states constructed in the intrinsic reference frame that defines nuclear deformations. In hadron physics, intrinsic quark-gluon structures can be revealed by means of generalized parton distributions ~\cite{(Bel05a),(Fra05),(Con21)}, and this technique has recently been extended to light nuclei such as    $^4$He~\cite{(Mar26)}.

Inspired by the methods proposed for artificial atoms~\cite{Yannouleas2000,Romanovsky2006,Li2006,Yannouleas2023}, a powerful way to define an intrinsic reference frame for a nuclear many-body wave function with good total angular momentum is to use the conditional probability of finding one nucleon at $\bm{r}_1$ with spin $\sigma_1$ and isospin $\tau_1$ given that another nucleon with spin $\sigma_{2}$ and isospin $\tau_{2}$ is located at $\bm{r}_{2}$. This probability is related to the two-body density of the many-body wave function and enables reconstruction of the particle distribution in the intrinsic frame, which is hidden from view in the angular momentum eigenstates when relying on information offered by  one-body densities. In this Letter, we follow this route.

Symmetry considerations and conditional probabilities are crucial when modeling the initial state in ultra-relativistic heavy-ion collisions~\cite{Dobaczewski2025}. In addition, it has been realized recently that the proper way of describing the internucleon correlations in the initial state is through two- and three-body densities computed for the $A$-body wave function~\cite{Dobaczewski2025,Giacalone2023,Ke2025,Giacalone2025,Blaizot2025,Duguet2025}.

Here, we define two-body conditional probabilities derived from nuclear wave functions and use them to introduce the intrinsic frame and multipole moments. This framework allows discussion of nuclear shapes for states with angular momentum $J\leq 1/2$. General considerations are illustrated by results from nuclear density functional theory (DFT) calculations for spherical, prolate, and oblate intrinsic states. We present applications to the light nuclei $^{16}$O and $^{20}$Ne, a choice motivated by recent experiments on ultra-relativistic heavy-ion collisions and their theoretical interpretation~\cite{Giacalone2025}.

\bigskip
{\it Method\label{sec:method}}---%
Our analysis is based on determining the two-body local density matrix ${\trho}(x_1;x_2)$,
\begin{align}
\label{eqm:nonlocal_densities0}
    {\trho}&(x_1;x_2)=\langle\Psi|\hat{a}^{\dag}_{x_1}\hat{a}^{\dag}_{x_2}\hat{a}_{x_2}\hat{a}_{x_1}|\Psi\rangle,
\end{align}
where the space-spin-isospin coordinates and integrals are:
\begin{align}
&x\equiv\{\bfr\sigma\tau\} \quad\mbox{and}\quad \intsum\rmd{}x\equiv\int\rmd^3\bfr\sum_{\sigma\tau},
\end{align}
with $\sigma=\uparrow,\downarrow$ for spin up or down and $\tau=n,p$ for neutron or proton. $|\Psi\rangle$ is an arbitrary
$A$-body, in general symmetry-broken, nuclear state that in the coordinate representation has the form
$\Psi(x_{1},x_{2},...,x_{A})\equiv\langle{x_{1},x_{2},...,x_{A}}|\Psi\rangle$.

The corresponding two-body local probability densities to find nucleons 1 and 2, with spins and isospins $\sigma_1\tau_1$ and $\sigma_2\tau_2$, respectively, at space points $\bfr_1$ and $\bfr_2$ are:
\begin{align}
\label{eqm:local_densities1}
    {\tprob}&(x_1;x_2)=\intsum\rmd{}x_{3}\cdots\rmd{}x_{A}|\Psi(x_{1},...,x_{A})|^{2}.
\end{align}

For $A$-body states that conserve the neutron and proton numbers, $N_n\equiv{}N$ and $N_p\equiv{}Z$, separately, as is the case in our study, we have,
\begin{align}
\label{eqm:np_numbers}
    \hat{N}_n|\Psi\rangle=N|\Psi\rangle, \quad
    \hat{N}_p|\Psi\rangle=Z|\Psi\rangle,
\end{align}
where the neutron and proton particle-number operators are $\hat{N}_\tau=\int\rmd^3\bm{r}\sum_\sigma\hat{a}^{\dag}_{\bfr\sigma\tau}\hat{a}_{\bfr\sigma\tau}$. The two-body probability densities (\ref{eqm:local_densities1}) are normalized according to
\begin{align}
\label{eqm:normalized1}
{\tprob}(\bfr_1\sigma_1\tau;\bfr_2\sigma_2\tau )&=\tfrac{1}{N_\tau (N_      \tau-1) }{\trho}(\bfr_1\sigma_1\tau;\bfr_2\sigma_2     \tau ), \\
\label{eqm:normalized2}
{\tprob}(\bfr_1\sigma_1\BAR{\tau};\bfr_2\sigma_2\tau)&=\tfrac{1}{N_{\BAR{\tau}}N_\tau}{\trho}(\bfr_1\sigma_1\BAR{\tau};\bfr_2\sigma_2\tau),
\end{align}
where the tilde denote the reversed isospin index, $\BAR{n}=p$ and $\BAR{p}=n$.

Having Eqs.~(\ref{eqm:normalized1}) and~(\ref{eqm:normalized2}) in mind, we now proceed by considering the two-body local density matrix (\ref{eqm:nonlocal_densities0}) and probability densities (\ref{eqm:local_densities1}) determined for $A$-body wave function $\Psi_{JM}(x_{1},...,x_{A})\equiv\langle{x_{1},x_{2},...,x_{A}}|\Psi_{JM}\rangle$ that conserves the total angular momentum $J$ and its projection $M$:
\begin{align}
\label{eqm:local_densities23}
    {\trho}^{J'M'}_{JM}&(x_1;x_2)=\langle\Psi_{J'M'}|\hat{a}^{\dag}_{x_1}\hat{a}^{\dag}_{x_2}\hat{a}_{x_2}\hat{a}_{x_1}|\Psi_{JM}\rangle,
    \\
\label{eqm:local_densities22}
    {\tprob}^{J'M'}_{JM}&(x_1;x_2)=\intsum\rmd{}x_{3}\cdots\rmd{}x_{A}\Psi_{JM}(x_{1},...,x_{A})
    \times \nonumber \\ &\hspace*{38mm}
    \Psi^*_{J'M'}(x_{1},...,x_{A}).
\end{align}
Both ${\trho}$ and ${\tprob}$ inherit the angular-momentum quantum numbers and become rotational bi-tensors. Anticipating the angular-momentum-symmetry restoration, we keep the total angular momenta, $J$ and $J'$, distinct so that we can treat the transition densities on equal footing. Nevertheless, even for the diagonal $J=J'$ densities, the magnetic quantum numbers, $M$ and $M'$ must be kept different, because they define the polarisation substates of $|\Psi_{JM}\rangle$. The density matrix (\ref{eqm:local_densities23}) can be related to observables when the polarisation amplitudes $a_{JM}$ are known in a given experiment, which is then performed for the state $\sum_M{}a_{JM}|\Psi_{JM}\rangle$.

Details on deriving the spectroscopic density matrix (\ref{eqm:local_densities23}) for angular-momentum symmetry-restored states are provided in End Matter. After establishing the two-body density matrix ${\trho}_0(x_1;x_2)\equiv{\trho}_{00}^{00}(x_1;x_2)$ of the $J=0$ symmetry-restored state $|\Psi_{00}\rangle$, we have at our disposal the corresponding two-body local density probabilities given by Eqs.~(\ref{eqm:normalized1}) and (\ref{eqm:normalized2}), namely ${\tprob}_0(\bfr_1\sigma_1\tau;\bfr_2\sigma_2\tau)$ and ${\tprob}_0(\bfr_1\sigma_1\BAR{\tau};\bfr_2\sigma_2\tau)$.

We must emphasize that the two-body local density probabilities for $J=0$ states are scalar functions of the space-spin positions, that is, they are invariant under {\it simultaneous} space rotations of the variables $\bm{r}_1\sigma_1$ and $\bm{r}_2\sigma_2$. In other words, information about the system is encoded only in the {\it relative} positions of particles 1 and 2. For further analysis, we therefore fix the position of particle 2 on the $z$-axis at a distance $z_2$ from the center of mass.

The resulting two-body local density probabilities remain invariant under rotations about the $z$-axis. That is, rotationally invariant two-body local density probabilities capture only axial information about the system~\footnote{As specified in Ref.~\cite{Dobaczewski2025} and exploited in a subsequent publication~\cite{(Meh26)}, extracting information about triaxiality requires accounting for three-body probabilities.}. The meaningful information is thus contained in the two-body local density probabilities ${\tprob}_0(r_{\perp1}z_1\sigma_1\tau_1;z_2\sigma_2\tau_2)$, where $r_{\perp}$ is the cylindrical coordinate in the plane perpendicular to the $z$-axis.

At this point, we are ready to define the essential element of the method, namely the conditional probability, which relates the two-body local density probability to a specific interpretation. Since finding a nucleon at the space-spin-isospin position $x_1$ and finding one at $x_2$ are independent probabilistic events, we can meaningfully ask the question, what is the probability to find it at $x_1$ {\it under the condition} that there is one at $x_2$. This defines the conditional probability
\begin{align}
\label{conditional_probability}
{\tcond}_0(x_1|x_2)
&=\frac{{\tprob}_0(x_1;x_2)}{\intsumtext\rmd{}x{\tprob}_0(x;x_2)}
=\frac{{\tprob}_0(x_1;x_2)}{{\oprob}_0(x_2)},
\end{align}
where ${\oprob}_0(x_2)$ is the probability density to find a nucleon with spin $\sigma_2$ and isospin $\tau_2$ at space point $\bfr_2$. We note that the conditional probability is for any $x_2$  normalized as $\intsumtext\rmd{}x_1\,{\tcond}_0(x_1|x_2)=1$.

For the purpose of analyzing the conditional probabilities determined in this Letter, we change their normalization, average their spins, and express them in terms of the two-body local density matrices, \disregard{%
for like and unlike isospins,
\begin{align}
\label{conditional_probability_like}
{\bar{\tcond}}^{\tau\tau}_0(\bfr_1|\bfr_2\sigma_2)
&\equiv{}(N_\tau-1)\sum_{\sigma_1} {\tcond}_0(\bfr_1\sigma_1\tau|\bfr_2\sigma_2\tau)
\nonumber \\ &
=\frac{\sum_{\sigma_1}{\trho}_0(\bfr_1\sigma_1\tau;\bfr_2\sigma_2\tau)}{{\rho}_0^\tau(\bfr_2)/2},
\\
\label{conditional_probability_unlike}
{\bar{\tcond}}^{\BAR{\tau}\tau}_0(\bfr_1|\bfr_2\sigma_2)
&\equiv{}N_{\BAR{\tau}}\sum_{\sigma_1} {\tcond}_0(\bfr_1\sigma_1\BAR{\tau}|\bfr_2\sigma_2\tau)
\nonumber \\ &
=\frac{\sum_{\sigma_1}{\trho}_0(\bfr_1\sigma_1\BAR{\tau};\bfr_2\sigma_2\tau)}{{\rho}_0^\tau(\bfr_2)/2},
\end{align}
}
\begin{align}
\label{conditional_probability2}
{\bar{\tcond}}^{\tau_1\tau_2}_0(\bfr_1|\bfr_2\sigma_2)
&\equiv{}F_{\tau_1\tau_2}\sum_{\sigma_1} {\tcond}_0(x_1|x_2)
=\frac{\sum_{\sigma_1}{\trho}_0(x_1;x_2)}{{\rho}_0^{\tau_2}(\bfr_2)/2},
\end{align}
where ${\rho}_0^\tau(\bfr_2)=N_\tau\sum_{\sigma_2}\oprob_0(\bfr_2\sigma_2\tau)$ are the one-body local densities of protons or neutrons, $F_{\tau\tau}=N_{\tau}-1$, and $F_{\BAR{\tau}\tau}=N_{\BAR{\tau}}$.

The conditional densities (\ref{conditional_probability}) and (\ref{conditional_probability2}) are called tomographic, as they express probability or particle density distributions as functions of $\bfr_1$, provided the observer point is located at $\bfr_2$. This is analogous to medical tomographic imaging, where the object's cross-section is shown as a function of the observation plane's position.

The background formalism presented above applies to any nuclear state $|\Psi\rangle$ and can be implemented in various many-body approaches. In what follows, we present examples obtained within nuclear DFT~\cite{(Sch19b)}.

\begin{figure}[t]
\centering\includegraphics[width=0.48\textwidth]{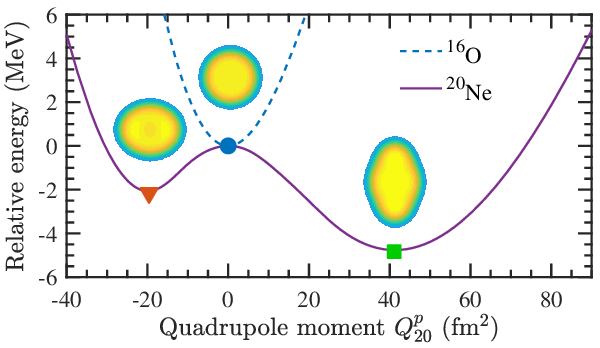}
\caption{Constrained HF energies of $^{16}$O and $^{20}$Ne relative to those obtained at the intrinsic one-body proton quadrupole moment $Q^p_{20}=0$.
\label{fig_PES}}
\end{figure}

\begin{figure*}[t]
\textbf{}\centering\includegraphics[width=0.99\textwidth]{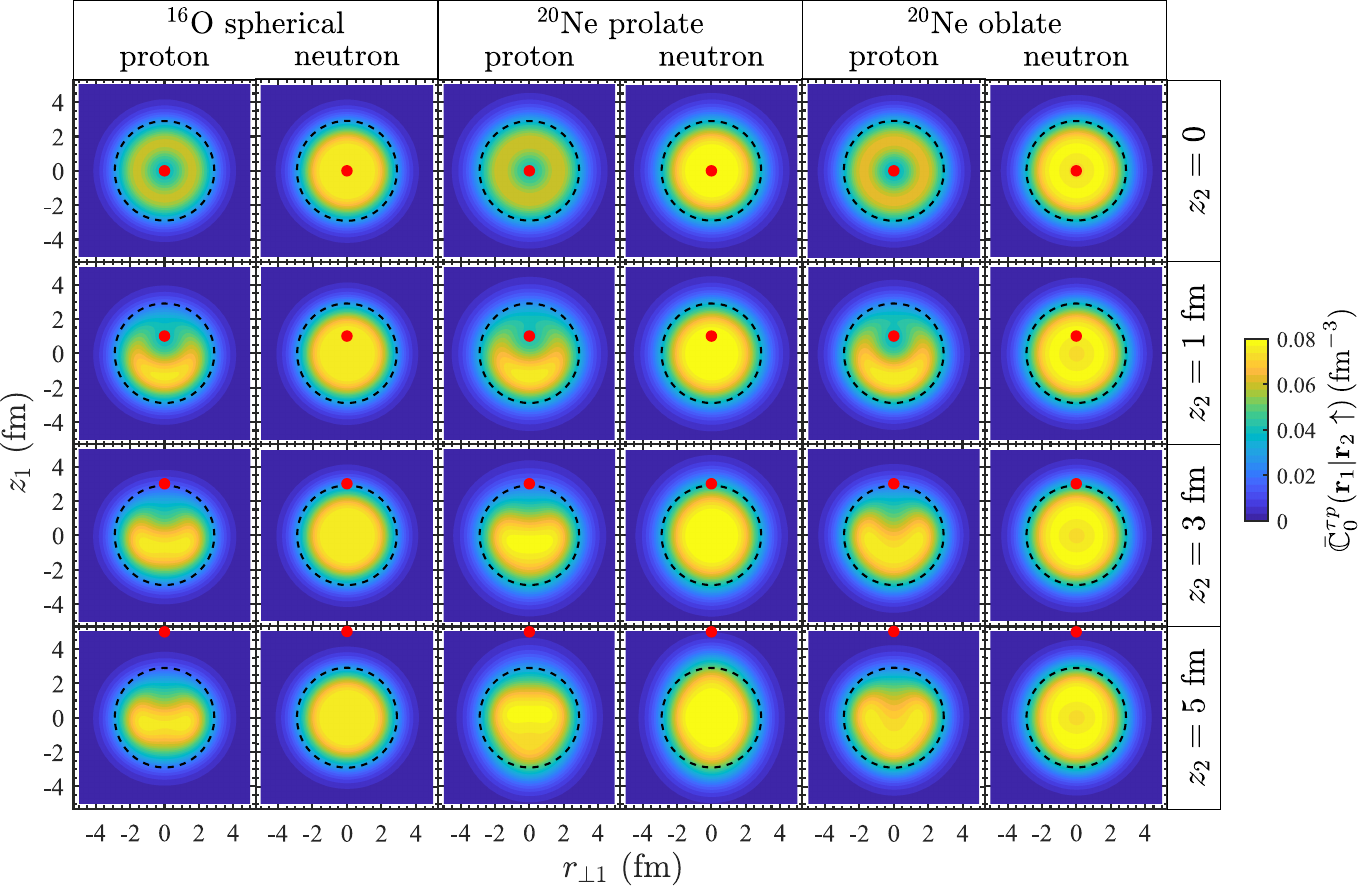}
\caption{Proton and neutron conditional densities ${\bar{\tcond}}^{\tau{}p}_0(r_{\perp1}z_1|z_2\uparrow)$,  Eq.~(\protect\ref{conditional_probability2}), plotted in the $r_{\perp1}$--$z_1$ plane for the $0^+$ states of $^{16}$O (spherical minimum), and $^{20}$Ne (prolate and oblate minima). The proton observer's positions are located at $z=z_2$ and are denoted by the red dots. To guide the eye, dashed circles are drawn in all panels at $r=3$\,fm.\label{fig_CPD6}}
\end{figure*}
{\it Results}---%
We illustrate the main features of the method by analyzing the results for $^{16}$O and $^{20}$Ne, obtained using code {\sc hfodd} (v3.34e)~\cite{(Dob21g),(Dob26a)} with Skyrme UNEDF1 functional~\cite{(Kor10c)} without pairing~\footnote{For technical reasons related to the implementation in the code, we performed the $^{20}$Ne calculations in the BCS pairing mode with fixed, small pairing gaps of $\Delta_n=\Delta_p=50$\,keV, which, for all practical purposes, is equivalent to no pairing.}.  In Fig.~\ref{fig_PES}, we present constrained HF energies for these two nuclei as functions of the intrinsic axial proton quadrupole moment $Q^p_{20}$. For the doubly magic nucleus $^{16}$O, the energy minimum occurs at spherical shape,  $Q^p_{20}=0$. For the open-shell nucleus $^{20}$Ne, the global minimum is found at $Q^p_{20}=+41.1$\,fm$^2$, corresponding to an elongated (prolate) shape, and an excited flattened (oblate) configuration at $Q^p_{20}=-19.6$\,fm$^2$. In the following, we present results obtained by projecting the $J=0$ angular momentum component from the HF wave functions corresponding to these three representative nuclear shapes.

Figure~\ref{fig_CPD6} presents the tomographic proton ${\bar{\tcond}}^{pp}_0(\bfr_1|{\bfr_2}\uparrow)$ and neutron ${\bar{\tcond}}^{np}_0(\bfr_1|\bfr_2\uparrow)$ densities, Eq.~(\ref{conditional_probability2}), with the proton spin-up observer located at $\bfr_2=(0,0,z_2)$. The color maps show the axially-symmetric densities in the ($r_{\perp1}, z_1$) plane.

For the observer located at $z_2=0$ (top panels), the tomographic densities are spherical, even for the deformed configurations of $^{20}$Ne. This reflects the fact that all corresponding symmetry-restored $J=0$ states are spherically symmetric. However, as soon as the proton observer shifts away from the center, the proton tomographic densities become deformed and oblate. The principal cause of this effect is the proton spin-up hole that appears around the observer due to the Pauli exclusion principle. This Pauli hole is present at all values of $z_2$.
The neutron conditional density ${\bar{\tcond}}^{np}_0(\bfr_1|\bfr_2\uparrow)$ is not affected by the proton Pauli effect, and hence it is of particular interest for shape imaging. Consequently, for all proton observer positions, the neutron tomographic densities of $^{16}$O are spherical. For the prolate configuration of $^{20}$Ne, the conditional density ${\bar{\tcond}}^{np}_0$
develops appreciable prolate shape at large values of $z_2$.

\begin{figure*}[tbh]
\centering\includegraphics[width=0.98\textwidth]{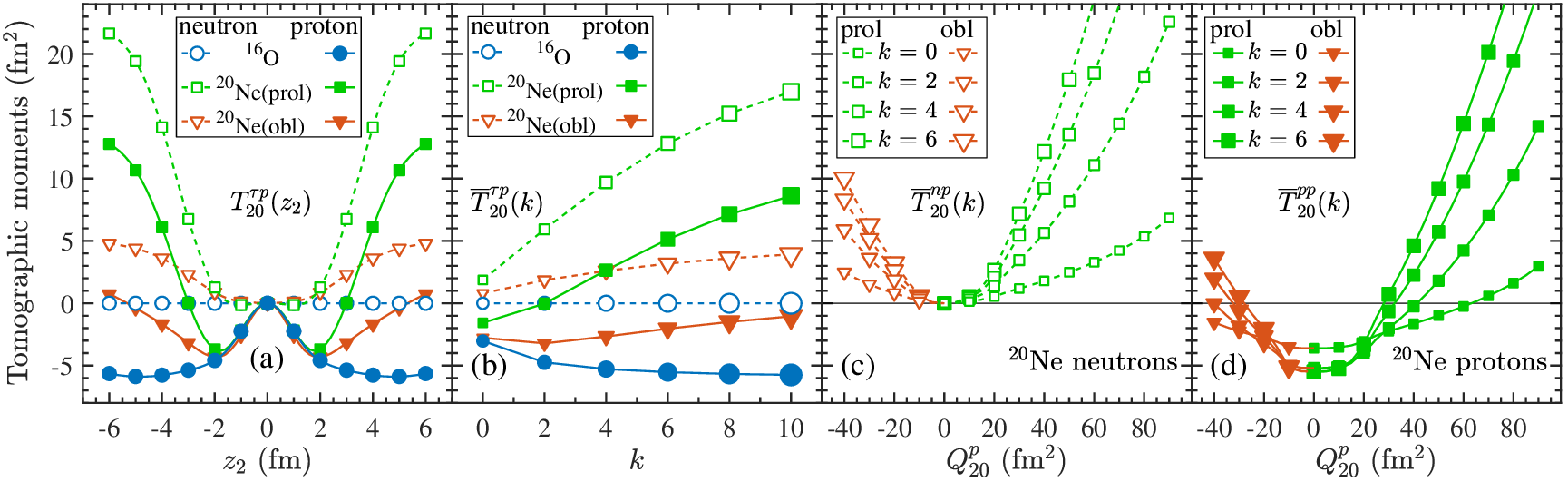}
\caption{(a) Neutron (open symbols) and proton (full symbols) quadrupole tomographic moments $T^{\tau{}p}_{20}(z_2)$, Eq.~(\protect\ref{tomographic_moment}), for the energy minima, Fig.~\ref{fig_PES}, of $^{16}$O (circles) and $^{20}$Ne (squares and triangles) as functions of the proton spin-up observer's position $z_2$. (b) Same as in (a), but for the cumulative averages $\overline{T}^{\tau{}p}_{20}(k)$ (\protect\ref{cumulative_tomographic_moment}) as functions of $k$, with symbol sizes indicating the $k$-values. (c) Same as in (b) but for neutron cumulative averages as functions of the intrinsic one-body proton quadrupole moment $Q^p_{20}$. (d) Same as in  (c) but for the proton cumulative averages.\label{Tomography.Combined}}
\end{figure*}
It is interesting to compare the results obtained for the spherical, prolate, and oblate minima. In Fig.~\ref{Tomography.Combined}(a), we show the corresponding neutron and proton quadrupole ($\lambda=2$) tomographic moments,
\begin{align}
\label{tomographic_moment}
T^{\tau{}p}_{\lambda0}(z_2)=
\int\rmd^3\bfr_1\,Q_{\lambda0}(\bfr_1){\bar{\tcond}}^{\tau{}p}_0(r_{\perp1}z_1|z_2\uparrow),
\end{align}
as functions of the proton spin-up observer's position $z_2$. Again, we emphasize that the axial symmetry of the conditional probability ${\bar{\tcond}}^{\tau{}p}_0(r_{\perp1}z_1|z_2\uparrow)$ implies that no information can be obtained on the non-axial multipole tomographic moments $T^{\tau{}p}_{\lambda\mu}(z_2)$ for $\mu\neq0$.

The Pauli hole appears to dominate the shapes, with the proton tomographic densities becoming oblate at $|z_2|\lesssim2$\,fm and indistinguishable across the spherical, prolate, and oblate intrinsic minima. At the same time, the neutron tomographic densities stay almost spherical. Only at the surface and peripheral observer positions do the quadrupole tomographic moments gradually reach positive values, with the neutron moments exceeding proton moments. The quadrupole tomographic neutron moments computed at oblate intrinsic minima appear to be positive, but they are significantly smaller than those at the prolate minima. It thus appears that differentiation between the prolate and oblate intrinsic shapes can be challenging based on the knowledge of $T^{np}_{20}(z_2)$. In heavy nuclei, which we study in a forthcoming publication, we expect the Pauli hole's role to be relatively weaker.

To get a sense of the overall magnitude of quadrupole tomographic moments, we consider their cumulative averages weighted by the moments of particle densities at the observer positions $z_2$~\footnote{We remark that because  ${\rho}_0(\bfr)$
is a rotational scalar, the volume integrals must vanish  for $\lambda>0$, that is, $\int\rmd^3\bfr\,T^{\tau{}\tau'}_{\lambda0}(\bfr)\,r^k\,{\rho}_0^{\tau'}(\bfr)=0$.}:
\begin{align}
\label{cumulative_tomographic_moment}
\overline{T}^{\tau{}\tau'}_{\lambda0}(k)=\frac{\int_0^\infty\rmd z_2 T^{\tau{}\tau'}_{\lambda0}(z_2)\,z^k_2\,{\rho}_0^{\tau'}(z_2)}
{\int_0^\infty\rmd z_2\,z^k_2\, {\rho}_0^{\tau'}(z_2)},
\end{align}
with higher orders $k$ probing the quadrupole tomographic moments at larger observer distances.

Figure~\ref{Tomography.Combined}(b) shows the cumulative quadrupole moments calculated up to order $k=10$. We observe that, at every order, the values of $\overline{T}^{\tau{}\tau'}_{20}(k)$ differ significantly from the intrinsic moments shown in Fig.~\ref{fig_PES}. The negative proton moments of about $-$6\,fm$^2$ obtained for the spherical state of $^{16}$O illustrate the magnitude of the Pauli hole's effect.

Finally, in Figs.~\ref{Tomography.Combined}(c) and (d), we show the neutron and proton cumulative quadrupole moments, respectively, as functions of the intrinsic one-body proton quadrupole moment $Q^p_{20}$ in $^{20}$Ne. It is remarkable that, as the intrinsic oblate moment increases, both neutron and proton tomographic moments are positive. This again shows that the two-body density characteristics may not directly correspond to intrinsic shapes. On the other hand, at large  (positive or negative) values of $Q^p_{20}$, tomographic moments are large.

\bigskip
{\it Conclusions}---%
We presented the application of nuclear two-body conditional probabilities to characterize the intrinsic multipole shapes of nuclei in good-angular-momentum states. To this end, we introduced tomographic multipole moments that describe matter distributions as observed from the fixed position of a single nucleon. Our results showed that the tomographic moments $T^{pp}$ and $T^{nn}$  do not distinguish well among spherical, prolate, and oblate intrinsic states, are strongly affected by Pauli repulsion, and differ appreciably from intrinsic one-body moments. The Pauli hole is not present in $T^{np}$ and $T^{pn}$, and hence these moments are better indicators of nuclear shapes: they vanish at spherical shapes, and they are large for configurations with large intrinsic deformations.  This demonstrates that the tomographic characteristics can distinguish between deformed and spherical intrinsic shapes.

The tomographic multipole moments thus constitute independent quantum observables related to nuclear shapes. Since these quantities explicitly depend on the relative positions of nucleons within the nucleus, they constitute better characteristics of the initial state formed in ultra-relativistic heavy-ion collisions than one-body multipole moments, which often cannot be defined within symmetry-conserved theories~\cite{Dobaczewski2025}.

\bigskip
\begin{acknowledgments}
This work was partially supported by STFC Grant Nos.~ST/V001035/1 and~ST/Y000285/1 and by a Leverhulme Trust Research Project Grant.
This material is based upon work supported by the U.S. Department of Energy under Award Number DE-SC0013365  (Office of Science, Office of Nuclear Physics), DE-SC0023175 (Office of Science, NUCLEI SciDAC-5 collaboration).
We thank the CSC-IT Center for Science Ltd., Finland, and the IFT Computer Center at the University of Warsaw, Poland, for the allocation of computational resources.
This project was partly undertaken on the Viking Cluster, a high-performance compute facility provided by the University of York. We are grateful for computational support from the University of York High Performance Computing service, Viking, and the Research Computing team.
We thank Grammarly\textsuperscript{\textregistered} for its support with English writing.
\end{acknowledgments}

{\it Data availability}---The data that support the findings of this article are openly available~\cite{rep-Tomography}.


%

\clearpage

\centerline{{\bf End Matter}}\bigskip

{\it Two-body local density}---%
The symmetry-conserving state obtained by restoring the angular-momentum symmetry can be written as \cite{(She21a)}:
\begin{align}
\label{eqm:imk}
|\Psi_{JM}\rangle &= \sum_K g_K \hat{P}^J_{MK} |\Psi\rangle \nonumber \\
&= \tfrac{2J+1}{8\pi^2}\int\rmd\Omega\, \sum_K g_K D^{J*}_{MK}(\Omega)\;
\hat{R}(\Omega)|\Psi\rangle,
\end{align}
where $\hat{R}(\Omega)$ is the operator that rotates the wave function by three Euler angles $\Omega$, and $g_K$ are the $K$-mixing amplitudes such that the states $|\Psi_{JM}\rangle$ are normalized $\langle\Psi_{J'M'}|\Psi_{JM}\rangle= \delta_{J'J}\delta_{M'M}$~\cite{(Rin80)}:
\begin{align}
\label{eqm:norm}
\sum_{K'K} g^*_{K'}g_K \langle\Psi|\hat{P}^{J}_{K'K}|\Psi\rangle =1.
\end{align}

The two-body local density matrix (\ref{eqm:local_densities23}) of the symmetry-restored state (\ref{eqm:imk}) becomes:
\begin{align}
\label{eqm:AMP_density_matrix}
    &{\trho}_{JM}^{J'M'}(x_1;x_2)=
    \left(\tfrac{2J+1}{8\pi^2}\right)^2\int\int\rmd\Omega\rmd\Omega'\, \sum_{K'K} g^*_{K'} g_K
\times \nonumber \\
&D^{J'}_{M'K'}(\Omega') D^{J*}_{MK}(\Omega)\;
\langle\Psi|\hat{R}^\dagger(\Omega')\hat{a}^{\dag}_{x_1}\hat{a}^{\dag}_{x_2}\hat{a}_{x_2}\hat{a}_{x_1}
\hat{R}(\Omega)|\Psi\rangle.
\end{align}
We note that the double integral over the Euler angles $\Omega$ and $\Omega'$ splits into two  integrals:
\begin{align}
\label{eqm:AMP_density_matrix_split1}
    {\htrho}^{J'M'K'}_{JM{\Kbis}}&(x_1;x_2)=
    \tfrac{2J+1}{8\pi^2}\int\rmd\Omega'\, D^{J*}_{M{\Kbis}}(\Omega') D^{J'}_{M'K'}(\Omega')
\times \nonumber \\
&\hat{R}^\dagger(\Omega')\hat{a}^{\dag}_{x_1}\hat{a}^{\dag}_{x_2}\hat{a}_{x_2}\hat{a}_{x_1}
\hat{R}(\Omega'), \\
\label{eqm:AMP_density_matrix_split2}
    {\trho}_{JM}^{J'M'}&(x_1;x_2)=
    \tfrac{2J+1}{8\pi^2}\int\rmd\Omega''\, \sum_{K'K{\Kbis}}  g^*_{K'}g_K
\times \nonumber \\
& D^{J*}_{{\Kbis}K}(\Omega'')
\langle\Psi|{\htrho}^{J'M'K'}_{JM{\Kbis}}(x_1;x_2)
\hat{R}(\Omega'')|\Psi\rangle,
\nonumber \\
    = \sum_{K'K{\Kbis}} & g_K g^*_{K'} \;
\langle\Psi|{\htrho}^{J'M'K'}_{JM{\Kbis}}(x_1;x_2)
 \hat{P}^{J}_{{\Kbis}K}|\Psi\rangle,
\end{align}
where we changed the integration variables as $\Omega=\Omega'\cdot\Omega''$, that is, $\hat{R}(\Omega)=\hat{R}(\Omega')\hat{R}(\Omega'')$, and used Eq.~4.7.1(1) of Ref.~\cite{(Var88)}.

In Eq.~(\ref{eqm:AMP_density_matrix_split1}), the integration over the Euler angles can be performed without any reference to state $|\Psi\rangle$ or to the transition density matrix needed in Eq.~(\ref{eqm:AMP_density_matrix_split2}). To determine the bi-tensor density operator~(\ref{eqm:AMP_density_matrix_split1}), we express the field operators $a_{x}$ in a suitable single-particle basis $\phi_{k}(x)$:
\begin{align}
\label{eqm:field_operators1}
a_{x}&=\sum_{k}\phi_{k}(x)a_{k}.
\end{align}
The active rotations of operators in Eq.~(\ref{eqm:AMP_density_matrix_split1}) can be transformed into passive rotations of the space-spin arguments:
\begin{align}
\label{eqm:field-operators2}
\hat{R}^\dagger(\Omega)a_{x}
\hat{R}(\Omega)&=\sum_{k}\phi^{\Omega}_{k}(x)a_{k}, \quad\mbox{for}
\\
\label{eqm:field-operators3}
\phi^{\Omega}_{k}(\bfr\sigma\tau)&=\phi_{k}\left(\hat{R}(\Omega)\bfr\sigma\hat{R}^\dagger(\Omega)\tau\right),
\end{align}
where $\hat{R}(\Omega)\bfr\sigma\hat{R}^\dagger(\Omega)$ denotes the space-spin point $\bfr\sigma$ inverse rotated by the angles $\Omega$.

The density operators~(\ref{eqm:AMP_density_matrix_split1})
can now be expressed in terms of standard two-body operators:
\begin{align}
\label{eqm:AMP_density_matrix_split3}
    {\htrho}^{{\LAMBDA}{\MU}{\NU}}(x_1;x_2)&=
    \sum_{\substack{k'_1k'_2\\k_1k_2}}{\trho}^{{\LAMBDA}{\MU}{\NU}}_{k'_1k'_2;k_1k_2}(x_1;x_2)
    a^\dagger_{k'_1} a^\dagger_{k'_2}a_{k_2}a_{k_1},
\end{align}
with
\begin{align}
\label{eqm:AMP_density_matrix_split4}
    {\trho}^{{\LAMBDA}{\MU}{\NU}}_{k'_1k'_2;k_1k_2}(x_1;x_2)&=
    \tfrac{2J+1}{8\pi^2}\int\rmd\Omega\, D^{{\LAMBDA}}_{{\MU}{\NU}}(\Omega)
    \times \nonumber \\
&\phi^{\Omega*}_{k'_1}(x_1)\phi^{\Omega*}_{k'_2}(x_2)\phi^{\Omega}_{k_1}(x_1)\phi^{\Omega}_{k_2}(x_2),
\end{align}
which yields
\begin{align}
\label{eqm:AMP_density_matrix_split5}
    {\htrho}^{J'M'K'}_{JM{\Kbis}}(x_1;x_2)=&
    (-1)^{{\Kbis}-M}\sum_{{\LAMBDA}{\MU}{\NU}}C^{{\LAMBDA}{\MU}}_{J,-M;J'M'}
    \times \nonumber \\
    &C^{{\LAMBDA}{\NU}}_{J,-{\Kbis};J'K'}
    {\htrho}^{{\LAMBDA}{\MU}{\NU}}(x_1;x_2),
\end{align}
and finally,
\begin{align}
\label{eqm:AMP_density_matrix_split6}
    {\trho}_{JM}^{J'M'}&(x_1;x_2)=
     \!\!\sum_{K'K{\Kbis}} \!\!g_K g^*_{K'}
     (-1)^{{\Kbis}-M}\!\!\!\!\sum_{{\LAMBDA}{\MU}{\NU}}\!\!\!\!C^{{\LAMBDA}{\MU}}_{J,-M;J'M'}
    \times \nonumber \\
     &C^{{\LAMBDA}{\NU}}_{J,-{\Kbis};J'K'}
\langle\Psi|{\htrho}^{{\LAMBDA}{\MU}{\NU}}(x_1;x_2)
 \hat{P}^{J}_{{\Kbis}K}|\Psi\rangle.
\end{align}
In Eq.~(\ref{eqm:AMP_density_matrix_split5}), we took advantage of the identity~\cite{(Var88)}:
\begin{align}
\label{eqm:AMP_density_matrix_tens2}
D^{J*}_{M{\Kbis}}(\Omega)D^{J'}_{M'K'}(\Omega)=&
(-1)^{{\Kbis}-M}\sum_{{\LAMBDA}{\MU}{\NU}}C^{{\LAMBDA}{\MU}}_{J,-M;J'M'}
    \times \nonumber \\ &
C^{{\LAMBDA}{\NU}}_{J,-{\Kbis};J'K'}D^{{\LAMBDA}}_{{\MU}{\NU}}(\Omega),
\end{align}
where the Clebsch-Gordan coefficients define the summation range over ${\LAMBDA}$ as $|J-J'|\leq{\LAMBDA}\leq{}J+J'$, and ${\LAMBDA}$, ${\MU}$, and ${\NU}$ always take integer values.

Equations (\ref{eqm:AMP_density_matrix_split3})--(\ref{eqm:AMP_density_matrix_split6}) can be reduced to the case of diagonal spectroscopic densities, $J=J'=M=M'$, evaluated for aligned states, $J=J'=K=K'$, which are essential in odd nuclei~\cite{(Dob26f)}. They are given by:
\begin{align}
\label{eqm:AMP_density_matrix_split33}
    &{\htrho}^{{\LAMBDA}0{\NU}}(x_1;x_2)=
    \!\!\!\sum_{k'_1k'_2;k_1k_2}\!\!\!{\trho}^{{\LAMBDA}0{\NU}}_{k'_1k'_2;k_1k_2}(x_1;x_2)
    a^\dagger_{k'_1} a^\dagger_{k'_2}a_{k_2}a_{k_1},
\end{align}
\begin{align}
\label{eqm:AMP_density_matrix_split34}
    {\trho}^{{\LAMBDA}0{\NU}}_{k'_1k'_2;k_1k_2}(x_1;x_2)&=
    \tfrac{2J+1}{8\pi^2}\int\rmd\Omega\, D^{{\LAMBDA}}_{0{\NU}}(\Omega)
    \times \nonumber \\
&\phi^{\Omega*}_{k'_1}(x_1)\phi^{\Omega*}_{k'_2}(x_2)\phi^{\Omega}_{k_1}(x_1)\phi^{\Omega}_{k_2}(x_2),
\end{align}
\begin{align}
\label{eqm:AMP_density_matrix_split35}
    {\htrho}^{JJJ}_{JJ{\Kbis}}(x_1;x_2)=&
    (-1)^{{\Kbis}-J}\sum_{{\LAMBDA}{\NU}}C^{{\LAMBDA}0}_{J,-J;JJ}
    \times \nonumber \\
    &C^{{\LAMBDA}{\NU}}_{J,-{\Kbis};JJ}
    {\htrho}^{{\LAMBDA}0{\NU}}(x_1;x_2),
\end{align}
\begin{align}
\label{eqm:AMP_density_matrix_split36}
    {\trho}_{J}(x_1;x_2)\equiv&{\trho}_{JJ}^{JJ}(x_1;x_2)=
     \sum_{{\LAMBDA}}C^{{\LAMBDA}0}_{J,-J;JJ}\sum_{{\Kbis}{\NU}}(-1)^{-\NU}\!\!
    \times \nonumber \\
     &
C^{{\LAMBDA}{\NU}}_{J,-{\Kbis};JJ}\frac{\langle\Psi|{\htrho}^{{\LAMBDA}0{\NU}}(x_1;x_2)
 \hat{P}^{J}_{{\Kbis}J}|\Psi\rangle}{\langle\Psi|
 \hat{P}^{J}_{JJ}|\Psi\rangle},
\end{align}
where, for better readability, we reduced the angular-momentum indices of the density matrices to ${\trho}_J(x_1;x_2)$.
If, in addition, the state $|\Psi\rangle$ is axial, the integrals over the Euler angles $\Omega$ can be reduced to a one-dimensional integral over the Euler angle $\beta$.

For the $J=0$ states that are important  in the context of this Letter, Eqs.~(\ref{eqm:AMP_density_matrix_split33})--(\ref{eqm:AMP_density_matrix_split36}) can be further reduced ($J=J'=M=M'=K=K'={\Kbis}={\LAMBDA}={\MU}={\NU}=0$):
\begin{align}
\label{eqm:AMP_density_matrix_split03}
&\hspace*{-3mm}
{\htrho}_0(x_1;x_2)=
    \!\!\!\!\sum_{k'_1k'_2;k_1k_2}\!\!\!\!{\trho}_{0;k'_1k'_2;k_1k_2}(x_1;x_2)
    a^\dagger_{k'_1} a^\dagger_{k'_2}a_{k_2}a_{k_1},
\\
\label{eqm:AMP_density_matrix_split04}
&\hspace*{-3mm}
{\trho}_{0;k'_1k'_2;k_1k_2}(x_1;x_2)=
    \tfrac{1}{8\pi^2}\int\rmd\Omega\,
\phi^{\Omega*}_{k'_1}(x_1)\phi^{\Omega*}_{k'_2}(x_2)
    \times \nonumber \\ &\hspace*{45mm}
\phi^{\Omega}_{k_1}(x_1)\phi^{\Omega}_{k_2}(x_2),
\\
\label{eqm:AMP_density_matrix_split06}
&\hspace*{-3mm}
{\trho}_0(x_1;x_2)\equiv{\trho}_{00}^{00}(x_1;x_2)=\frac{\langle\Psi|{\htrho}_0(x_1;x_2)\hat{P}^{0}_{00}|\Psi\rangle}
    {\langle\Psi|\hat{P}^{0}_{00}|\Psi\rangle},
\end{align}
where we further reduced the angular-momentum indices of the density matrices to a single index, that is, ${\htrho}_0(x_1;x_2)\equiv{\htrho}^{000}(x_1;x_2)$ and ${\trho}_{0;k'_1k'_2;k_1k_2}(x_1;x_2)\equiv{\trho}^{000}_{k'_1k'_2;k_1k_2}(x_1;x_2)$.

Equation~(\ref{eqm:AMP_density_matrix_split04}) defines the $J=0$ (monopole) matrix elements of the two-body local density operator~(\ref{eqm:AMP_density_matrix_split03}), which is then used in Eq.~(\ref{eqm:AMP_density_matrix_split06}) within the standard expression for the projection of the symmetry-breaking state $|\Psi\rangle$ onto $J=0$.

We note that Eqs.~(\ref{eqm:AMP_density_matrix_split03})--(\ref{eqm:AMP_density_matrix_split06}) are valid for an arbitrary $A$-body state $|\Psi\rangle$. For the HF states, discussed in this Letter, the transition matrix elements needed in Eq.~(\ref{eqm:AMP_density_matrix_split06}) can be evaluated using the standard generalized Wick theorem~\cite{(Rin80)}.

{\it One-body non-local density}---%
The one-body non-local density matrix $\rho(x;x')$ and probability densities $n(x;x')$ are  defined as follows,
\begin{align}
\label{eq:nonlocal_densities0}
    \rho&(x;x')=\langle\Psi|\hat{a}^{\dag}_{x'}\hat{a}_{x}|\Psi\rangle,
\\
\label{eq:local_densities1}
    n&(x;x')=\intsum\rmd{}x_{2}\cdots\rmd{}x_{A}\Psi(x,x_{2},...,x_{A})
\times \nonumber  \\ &\hspace*{36mm}
    \Psi^*(x',x_{2},...,x_{A}),
\end{align}
and, under conditions (\ref{eqm:np_numbers}), they are diagonal in isospin and normalized as:
\begin{align}
\label{eq:nonlocal_densities1}
n^\tau(\bfr\sigma;\bfr'\sigma')=\tfrac{1}{N_\tau}\rho^\tau(\bfr\sigma;\bfr'\sigma'),
\end{align}
for $n(\bfr\sigma\tau,\bfr'\sigma'\tau')=n^\tau(\bfr\sigma,\bfr'\sigma')\delta{\tau\tau'}$, $\rho(\bfr\sigma\tau,\bfr'\sigma'\tau')=\rho^\tau(\bfr\sigma,\bfr'\sigma')\delta{\tau\tau'}$. In particular, the two-body local probability density (\ref{eqm:local_densities1}) involves one-body local probability density $n(x)\equiv{}n(x;x)$ and one-body local density matrix $\rho(x)\equiv\rho(x;x)$ as
\begin{align}
\label{eq:local_densities9}
n(x)=\tfrac{1}{N_\tau}\rho(x)=\intsum\rmd{}x_{2}\tprob(x,x_2).
\end{align}



The one-body non-local density matrix (\ref{eq:nonlocal_densities0}) of the symmetry-restored state (\ref{eqm:imk}) can thus be written as:
\begin{align}
\label{eq:AMP_density_matrix_split3}
&\hat{\Rho}^{{\LAMBDA}{\MU}{\NU}}(x;x')=
    \sum_{k'k} \Rho^{{\LAMBDA}{\MU}{\NU}}_{k'k}(x;x')
    a^\dagger_{k'}a_{k},
\end{align}
\begin{align}
\label{eq:AMP_density_matrix_split4}
&\Rho^{{\LAMBDA}{\MU}{\NU}}_{k'k}(x;x')=
    \tfrac{2J+1}{8\pi^2}\int\rmd\Omega\, D^{{\LAMBDA}}_{{\MU}{\NU}}(\Omega)
\phi^{\Omega*}_{k'}(x')\phi^{\Omega}_{k}(x),
\end{align}
\begin{align}
\label{eq:AMP_density_matrix_split5}
&\hat{\rho}^{J'M'K'}_{JM{\Kbis}}(x;x')=
    (-1)^{{\Kbis}-M}\sum_{{\LAMBDA}{\MU}{\NU}}C^{{\LAMBDA}{\MU}}_{J,-M;J'M'}
\times \nonumber \\ &\hspace*{35mm}
C^{{\LAMBDA}{\NU}}_{J,-{\Kbis};J'K'}
    \hat{\Rho}^{{\LAMBDA}{\MU}{\NU}}(x;x'),
\end{align}
\begin{align}
\label{eq:AMP_density_matrix_split6}
\rho_{JM}^{J'M'}&(x;x')=
     \!\!\sum_{K'K{\Kbis}} \!\!g_K g^*_{K'}
     (-1)^{{\Kbis}-M}\!\!\!\!\sum_{{\LAMBDA}{\MU}{\NU}}\!\!\!\!C^{{\LAMBDA}{\MU}}_{J,-M;J'M'}
\times \nonumber \\ &\hspace*{10mm}
     C^{{\LAMBDA}{\NU}}_{J,-{\Kbis};J'K'}
\langle\Psi|\hat{\Rho}^{{\LAMBDA}{\MU}{\NU}}(x;x')
 \hat{P}^{J}_{{\Kbis}K}|\Psi\rangle.
\end{align}

The diagonal spectroscopic densities are:
\begin{align}
\label{eq:AMP_density_matrix_split13}
    \hat{\Rho}^{{\LAMBDA}0{\NU}}(x;x')&=
    \sum_{k'k} \Rho^{{\LAMBDA}0{\NU}}_{k'k}(x;x')
    a^\dagger_{k'}a_{k},
\\
\label{eq:AMP_density_matrix_split14}
    \Rho^{{\LAMBDA}0{\NU}}_{k'k}(x;x')&=
    \tfrac{2J+1}{8\pi^2}\int\rmd\Omega\, D^{{\LAMBDA}}_{0{\NU}}(\Omega)
\phi^{\Omega*}_{k'}(x')\phi^{\Omega}_{k}(x),
\\
\label{eq:AMP_density_matrix_split15}
    \hat{\rho}^{JJJ}_{JJ{\Kbis}}(x;x')&=
    \sum_{{\LAMBDA}{\NU}}(-1)^{-{\NU}}C^{{\LAMBDA}0}_{J,-J;JJ}
\times \nonumber \\ &\hspace*{15mm}
    C^{{\LAMBDA}{\NU}}_{J,-{\Kbis};JJ}
    \hat{\Rho}^{{\LAMBDA}0{\NU}}(x;x'),
\\
\label{eq:AMP_density_matrix_split16}
    \rho_J(x;x')\equiv&\rho_{JJ}^{JJ}(x;x')=
     \sum_{{\LAMBDA}}C^{{\LAMBDA}0}_{J,-J;JJ}\sum_{{\Kbis}{\NU}}(-1)^{-{\NU}}
\times \nonumber \\ &
C^{{\LAMBDA}{\NU}}_{J,-{\Kbis};JJ}\frac{\langle\Psi|\hat{\Rho}^{{\LAMBDA}0{\NU}}(x;x')
 \hat{P}^{J}_{{\Kbis}J}|\Psi\rangle}
{\langle\Psi|
 \hat{P}^{J}_{JJ}|\Psi\rangle}.
\end{align}

Finally, for the $J=0$ states, we have,
\begin{align}
\label{eq:AMP_density_matrix_split03}\hspace*{-3mm}
    \hat{\Rho}_0(x;x')&=
    \sum_{k'k} \Rho_{0;k'k}(x;x')
    a^\dagger_{k'}a_{k},
\\
\label{eq:AMP_density_matrix_split04}\hspace*{-3mm}
    \Rho_{0;k'k}(x;x')&=
    \tfrac{1}{8\pi^2}\int\rmd\Omega\,\phi^{\Omega*}_{k'}(x')\phi^{\Omega}_{k}(x),
\\
\label{eq:AMP_density_matrix_split06}\hspace*{-3mm}
    \rho_0(x;x')&=
\frac{\langle\Psi|\hat{\Rho}^{000}(x;x')
 \hat{P}^{0}_{00}|\Psi\rangle}
 {\langle\Psi|
 \hat{P}^{0}_{00}|\Psi\rangle},
\end{align}
where  $\hat{\Rho}_0(x;x')\equiv{\hat{\Rho}}^{000}(x;x')$, ${\rho}_{0;k'k}(x;x')\equiv{\rho}^{000}_{k'k}(x;x')$, and ${\rho}_0(x;x')\equiv{\rho}_{00}^{00}(x;x')$.

\end{document}